\documentclass[twocolumn, aps, amsfonts, amsmath, showpacs]{revtex4}

\usepackage{graphicx}

\def\hm{\hspace{1em}}

\begin{document}

\title{An dynamical-mean-field-theory investigation of specific heat and electronic structure of $\alpha$ and
$\delta$-plutonium}

\author{L. V. Pourovskii$^1$, G.~Kotliar$^2$, M. I. Katsnelson$^3$, and A. I. Lichtenstein$^4$}
\affiliation{
$^1$ Center de Physique Th\'{e}orique, \'{E}cole Polytechnique, 91128 Palaiseau Cedex, France \\
$^2$ Department of Physics and Astronomy and Center for Condensed Matter Theory, 
Rutgers University, Piscataway, New Jersey 08854-8019, USA \\
$^3$Institute for Molecules and Materials, Radboud University of Nijmegen,
NL-6525 ED Nijmegen, The Netherlands \\
$^4$Institut f\"ur Theoretische Physik, Universit\"at Hamburg,
20355 Hamburg, Germany}

\date{\today}

\begin{abstract}
We have  carried  out a comparative study
of the electronic specific heat and electronic structure of  $\alpha$ and
$\delta$-plutonium using dynamical mean field theory (DMFT).
We use the perturbative T-matrix and fluctuating exchange (T-matrix FLEX)  as a quantum impurity
solver. We considered two different physical pictures of plutonium.
In the first,  $5{f^5}+$, 
the perturbative treatment of electronic correlations has been carried out around 
the  non-magnetic (LDA) Hamiltonian, which results in an f occupation around a 
bit above   $n_f = 5 $.
In the second,   $5{f^6}-$, plutonium is viewed as being close to an  $5{f^6}$ configuration, and
perturbation theory is carried out around the  
(LDA+U)  starting point   
bit below    $n_f = 6 $.
In the latter case the  electronic 
specific heat coefficient $\gamma$ attains a smaller value in $\gamma$-Pu than in $\alpha$-Pu, in contradiction
to experiment, while in the former case our calculations reproduce the experimentally observed large increase of 
$\gamma$ in $\delta$-Pu as compared to the $\alpha$ phase. This enhancement  of  the electronic specific 
heat coefficient in 
$\delta$-Pu is due to strong electronic correlations present in this phase, which cause a substantial increase of 
the electronic effective mass, and high density of states at $E_F$. 
The densities of states of $\alpha$ and $\delta$-plutonium obtained  starting from  
the open-shell configuration are 
also in good agreement with the experimental photoemission spectra.
\end{abstract}

\maketitle

\section{Introduction}
Plutonium metal  is arguably the most complex elemental solid.
It exhibits a large number of allotropes, and in all the
phases 
its  Pauli-like magnetic 
susceptibility and resistivity are an order of magnitude larger than for simple metals. The phase transitions between
different phases are sometimes accompanied by very large discontinuities in volume
\cite{hand_act84}. 
There is a generalized suspicion that the   origin of the unique properties of Pu must
be  connected to  its unique  position in the actinide series lying  at the boundary between 
the early actinides where the f electrons are band-like and the late actinides where they 
are more localized and hence strongly correlated.
However there is currently no consensus on
the underlying electronic structure of Pu  and the precise mechanism  underlying  the anomalous behavior
of plutonium and the late actinides, 
and several efforts in developing realistic  approaches to this class of problems are currently being pursued.

The traditional electronic structure techniques have been unable to account for the unusual properties of Pu.
If one supposes itinerant nature of the Pu 5$f$ electrons,  {\it volume}  of both the   low-temperature $\alpha$ phase 
and that of the fcc high temperature $ \delta$ phase can be
described reasonably well \cite{sod97}. However, 
a significant spin polarization then results   in the LSDA calculations 
\cite{sod2001,kutep03,nikl03,pour_emto05,Solovyev:1991},
while a number 
experimental investigations have led to conclusion that Pu is non-magnetic (for review see \cite{Lash05}).

The strongly localized limit for the 5$f$ shell has been studied by means of the LDA+U technique 
\cite{savr2000,shick05,shor05,Bouchet:2000}.
Depending
on the type of the "double-counting" correction employed and  depending on the relative strength
of the multiplets and the spin orbit coupling and the  strength of the local 
Coulomb interaction between 5$f$ 
electrons one obtains either strongly magnetic $f^5$  or a  non-magnetic $f^6$ configurations for the localized $f$ shell
\cite{shick05,shor05}.
In the latter case one obtains the equilibrium volume of the $\delta$-phase in good agreement with the experiment
\cite{shick05}. That has led to conclusion that the ground state of $\delta$-Pu should be essentially in 
the $f^6$ non-magnetic 
configuration, with the 5$f$ shell being split due to the spin-orbit coupling into the completely 
filled $f^{5/2}$ and empty 
$f^{7/2}$ states, with the local Coulomb interaction $U$ further increasing this splitting. 
Therefore one predicts the 
ground state of $\delta$-Pu to be fairly similar to that of Am, though with substantially 
smaller degree of localization of the $f$ 
electrons. Similar results have been obtained recently by Shick et. al.~\cite{shick06} by means of LDA+DMFT using
Hubbard I as the impurity solver.

However, in this picture one expects to have almost no $f$-electrons in vicinity of the Fermi level, in
sharp contradiction to the experimental PES \cite{arko2000}, where the high peak attributed to $f$ states 
is clearly observed at the 
Fermi level. That deficiency has been addressed in Ref.~\cite{pour06}, where the dynamical fluctuations 
around $f^6$ ground state
have been added by means of the dynamical mean-field theory (DMFT) \cite{geo96} in conjunction with the perturbative
treatment of the local quantum fluctuations by the fluctuating exchange and T-matrix technique (T-matrix FLEX) 
\cite{bick89,kats02,pour05}. 
The dynamical fluctuations have  not changed the $f$ shell occupancy, however they led to essential modification of 
the density of states (DOS),  producing a peak which now is  closer  to ( but still separated by 
a finite energy from )    the Fermi level, 
substantially improving the  overall agreement between the theoretical DOS and experimental
PES for $\delta$-Pu.

Another school of thought, dating back to Johansson  \cite{Johansson:1975} 
suggests that  in the solid, plutonium is close to an $5f^5$ configuration.  
These ideas were supported by the first LDA+DMFT computations in this material by 
by  Savrasov {\it et al.} \cite{savr2001}.
At the reasonable value of the Coulomb interaction $U=$4 eV Savrasov {\it et al.} have
obtained a double-minima shape of the total energy versus volume curve, with positions of the minima corresponding
well to the experimental equilibrium volumes of the $\alpha$ and $\delta$ phases of Pu, respectively. The minima
have been assigned to the itinerant and localized state of the 5$f$ shell, thus the $\alpha \to \delta$ phase
transition is due to the Mott localization of the Pu $f$ electrons. The same explanation for the $\alpha \to \delta$ 
phase transition has been put forward by Katsnelson {\it et al.} ~\cite{kats92}.
The physical picture underlying this approach, is that the actinides provide a complex generalization of the localization-delocalization
DMFT phase diagram obtained in simple model Hamiltonians \cite{physics_today}, obtained by incorporating the realistic
band structure, the realistic electronic structure of the f electrons, and most important, by relaxing the position of the
atomic coordinates that were kept fixed in the model Hamiltonian approach.  
The LDA+DMFT DOS
reported in Ref.~\cite{savr2001} did not include realistic multiplets structure. Its recent incorporation
has lead to rather good agreement with the experimental PES, and in addition 
has shown that 
the fully  self consistent DMFT  solution is non-magnetic \cite{Shim:2006}, by allowing the possibility
of ordered states.
The ground state configuration
of both the $\alpha$ and $\delta$-Pu were  found to be close to the "open-shell" $f^5$ configuration, in the sense
that the f occupation in the ground state is slightly bigger than $n_f= 5$.


At this point we have,  within the Electronic Structure Method +DMFT
framework, two  rather different pictures of the ground state of Pu. In the "near $5f^6$"
of Refs.~\cite{shick05,shor05,pour06} one has essentially the $5f^6$ "closed-shell" non-magnetic configuration, 
(resulting from a strongly interacting one body Hamiltonian obtained in LDA+U)
with dynamical  fluctuations 
reducing the f occupancy by inducing virtual fluctuations into the $5f^5$ manifold,  ( and all other f occupancies).
Another 
approach of Ref.~\cite{savr2001} can be described 
as  starting from 
an  $f^5$ "open-shell" configuration, which is then 
screened by the spd  conduction electrons, what results in a Kondo resonance being formed at the Fermi level. 
The one electron Hamiltonian in this case is obtained from the non-magnetic weakly correlated LDA.
These two physical pictures  are not orthogonal.   In a solid the occupation is non integer,   and the theory
of the Anderson impurity model teaches us that at non integer occupancy the ground state of the model can
be adiabatically continued from  {\it both }  the $f^5 $ and $f^6 $  limits. 
Calculations based on both physical pictures produce  a  ground state 
and density of states of $\delta$-Pu in reasonable  agreement to experiment. 
Still, these two pictures are substantially different in content, and 
in order to chose which of two
approaches (and which one of the physical pictures) provides  the best description of the actual plutonium it is necessary to further compare  
their predictions for other physical quantities  sensitive to the  electronic structure. The  electronic 
specific heat coefficient $\gamma$ is a convenient choice as it is directly linked to strength of the electronic 
correlations. Moreover, it has been measured in both $\alpha$ and $\delta$-Pu, and the value of 
$\gamma$ in the $\delta$ phase (43  and 64 mJ K$^{-2}$ mol$^{-1}$ in $\delta$-Pu, stabilized by Ga and Al, respectively
 \cite{Lash03,Lash05}; in the range of 35-55 mJ K$^{-2}$ mol$^{-1}$ in the Pu$_{92}$Am$_{8}$ alloy \cite{havel06}) 
is
substantially larger than in $\alpha$-Pu (17 mJ K$^{-2}$ mol$^{-1}$ \cite{Lash03}).
The specific heat is a measure of the number of degrees of freedom available to the system at a given temperature.
It is much larger in the case of an open shell configuration screened by spd electrons, because the degrees of freedom
of the $f$ configuration are transferred to the low energy quasiparticle of the system.     
Hence, in the present work 
we compare and analyze 
predictions
of the two picture ( $f^5 + $ and $f^6 -$ ) implemented within LDA+DMFT
regarding values the $\gamma$ coefficient in the $\alpha$ and $\delta$-Pu.

\section{Computational method}

We have employed the LDA+DMFT technique on the basis of the full-potential linear MT-orbitals method (FPLMTO)
\cite{savr96} and in conjunction with the spin-orbital T-matrix FLEX quantum impurity solver \cite{pour05}. The
details of our technique are described in Ref.~\cite{pour05}. We start with the LDA or LDA+U calculations of
$\alpha$ and $\delta$-Pu by means of the FPLMTO method. The calculations of the $fcc$ $\delta-Pu$ phase 
were carried out at experimental lattice parameter   ($a=$4.64 \AA, $V$=25 \AA$^3$/at). The complex structure 
of the $\alpha$ phase has been mimicked by an orthorhombicaly distorted diamond structure (which has been proved
to have almost the same DOS and total energy as the actual $\alpha$-Pu structure \cite{bou04}) at experimental atomic
volume ($a=$3.63 \AA, $b/a$=1.61, $c/a$=2.09, $V$=20 \AA$^3$/at). We have employed 242 and 150 $k-$points in the 
irreducible Brillouin zone for the cubic and orthorhombic structures, respectively. In both the LDA+U and LDA+DMFT 
we have taken the values of 3 and 0.55 eV for the
parameters $U$ and $J$ of the local Coulomb interaction, respectively. In the LDA+U
calculations we have employed the "around mean-field" expression for the double counting term \cite{shick05}.

The one-particle Hamiltonian $H_t({\bf k})$  was obtained by 
L\"{o}wdin orthogonalization (using the square root of the overlap matrix) of the 
converged LDA or LDA+U calculations.
Below we designate as LDA+FLEX or LDA+U+FLEX the cases where the LDA or LDA+U
Hamiltonian, respectively, has been employed in the LDA+DMFT calculations in conjunction with the spin-orbit 
T-matrix FLEX
quantum impurity solver. First, the
local Green's function is obtained by means of the Brillouin zone (BZ)
integration
\begin{equation}\label{G_loc}
G(i\omega)=\sum_{\bf k} [(i\omega+\mu){\bf 1}-H_t({\bf
k})-\Sigma^{dc}(i\omega)]^{-1}
\end{equation}

where $\omega=(2n+1)\pi T$ are the fermionic Matsubara frequencies
for a given temperature $T$, $\mu$ is the chemical potential and
$\Sigma^{dc}(i\omega)$ is the local self-energy with a ``double
counting'' term subtracted, viewed as a functional of the local Greens
function $G(i\omega)$  which is solved for self consistently 
$\Sigma= \Sigma_{imp}- E_{dc}$.

Then within the DMFT scheme the local
self-energy is obtained by the solution of the many-body
problem for a single quantum impurity  coupled to an effective
electronic ``bath'' through the Weiss field function \cite{geo96}
\begin{equation}
{\mathcal G}^{-1}(i\omega)={G_{ff}}^{-1}(i\omega)+\Sigma^{dc}(i\omega).
\end{equation}

$G_{ff}$ denotes the ff block of $G$ in equation (\ref{G_loc})
In the spin-orbit T-matrix FLEX quantum impurity solver the local self-energy is obtained as a sum
of three contributions:
\begin{equation} \label{Sigma}
\Sigma=\Sigma^{(TH)}+\Sigma^{(TF)}+\Sigma^{(PH)},
\end{equation}
where $\Sigma^{(TH)}$ and $\Sigma^{(TF)}$ are the $T$-matrix
``Hartree'' and ``Fock'' contributions, respectively,
$\Sigma^{(PH)}$ is the particle-hole contribution. $\Sigma^{(TH)}$
and $\Sigma^{(TF)}$ are obtained by substitution in the
corresponding Hartree and Fock diagrams  the bare Coulomb
interaction with the frequency dependent $T$-matrix, the latter is
given by summation of the ladder diagrams in the particle-particle
channel. $\Sigma^{(PH)}$ is obtained by the RPA-type summation in
the particle-hole channel with the bare vertex being substituted
by static limit of the $T$-matrix \cite{kats02,pour05}. 

For the double counting term in  the DMFT computations  we have used the
static limit of the self energy  \cite{Lichtenstein:2001,kats02} for the LDA Hamiltonian and
the Hartree-Fork term for the LDA+U Hamiltonian \cite{pour05}. The
DMFT calculations have been carried out using the 512-points Matsubara frequencies corresponding
to temperature 950~K. We carried out DMTF iterations until
convergence in both the chemical potential $\mu$ and the local
self-energy was achieved. Then the Pade approximant was used
\cite{Vidb77} for analytical continuation of the local self-energy
to the real axis in order to obtain the DOS $n(E)=-1/\pi {\rm Tr}
\Im [G(E)]$. The electronic 
specific heat coefficient $\gamma$ was computed as $\frac{\pi^2}{3}k_B^2 tr [ N(E_F) Z^{-1}]$,
where $N(E_F)$ is the DOS matrix  at the Fermi level (which for rather low temperature used here can
be approximated to $\mu$) and $Z^{-1}$ is the quasiparticle residue matrix,  
 $Z^{-1}=I-\left. d \Im[\Sigma(E)]/dE\right|_{E=0}$.
where I is the identity matrix, and Sigma is  zero outside the f block.
The average mass enhancement  $\frac{m*}{m}$ is defined by  $ \frac{tr [Z^{-1} N_f ]}{tr [N_f]}$, where 
$N_f $  is the f  spectral function matrix at the Fermi level.

\section{Results}

\begin{figure}
\includegraphics[width=0.49\textwidth]{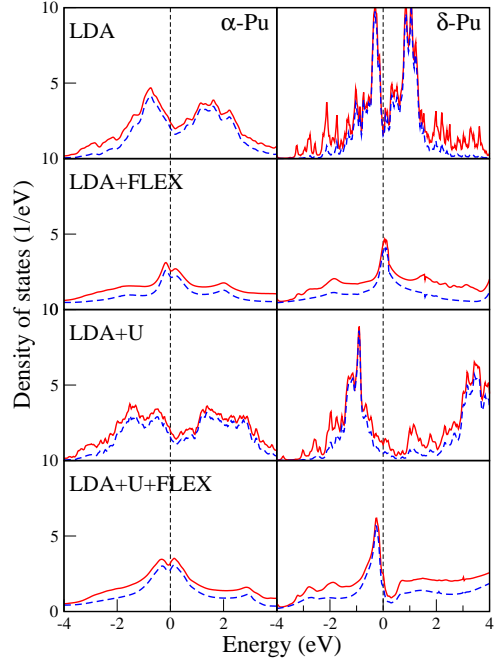}
\caption{The total (solid line) and partial $f$-states (dashed line) densities of states of $\alpha$ (left panel)
and $\delta$ (right panel) plutonium in the LDA, LDA+FLEX,
LDA+U and LDA+U+FLEX approaches, respectively.} \label{fig_dos}
\end{figure}

The calculated total and partial $f$-states LDA and LDA+U DOS are presented in Fig.~\ref{fig_dos}. Already in the
case of LDA one may clearly distinct the filled $f^{5/2}$ and empty $f^{7/2}$ states on the DOS of $\delta$-Pu. For
$\alpha$-Pu the separation between the $f^{5/2}$ and $f^{7/2}$ states is somewhat smeared out, as has been also observed
in Ref.~\cite{shor05}. However, in both the $\alpha$ and $\delta$ phases there is still substantial density of the 
$f$-states at the Fermi level $N_f(E_F)$, therefore within LDA one obtains an 
"open-shell" configuration of the $f$-band. 
The LDA+U calculations of $\delta$-Pu lead to familiar picture \cite{shick05,shor05,pour06} of non-magnetic  
configuration with the filled $f^{5/2}$ and empty $f^{7/2}$ states, well separated in energy by value of $U$, 
negligible $N_f(E_F)$ and the $f$ shell occupancy close to 6. 
In $\alpha$-Pu substantial broadening and intermixing of the $f^{5/2}$ and $f^{7/2}$ states 
persist in the LDA+U, retaining appreciable  $N_f(E_F)$.

The values of the electronic specific 
heat coefficient $\gamma$ and effective mass $m^*$ calculated within the LDA+FLEX and LDA+U+FLEX techniques  
 are listed in Table~\ref{table_gam}. One may notice, that
moderate values of the effective mass $m^*$ in the case of LDA+U+FLEX indicate rather weak correlations
present in the system. In fact, the effective mass $m^*$ in $\delta$-Pu appears to be somewhat smaller than
in $\alpha$-Pu. Similarly, in LDA+U+FLEX the $\gamma$ coefficient of specific heat in 
the $\delta$ phase is nearly two times smaller  as compared to $\alpha$-Pu, 
contrary to experimental measurements, which shows $\gamma$ bigger than two times and sometimes  
almost three  times {\it larger} in $\delta$-Pu. 
Almost order
of magnitude difference between experimental and theoretical $\gamma$ in the case of $\delta$-Pu means that
in the "closed-shell" configuration of $\delta$-Pu electronic correlations appear to be much weaker than one would
expect on basis of experimental evidences. This should be contrasted with  the LDA+FLEX calculations  which 
predict effective mass in $\delta$-Pu
to be quite large (4.02) and somewhat larger, than in $\alpha$-Pu. The LDA+FLEX technique also reproduce 
well experimental tendency of enhancement of the
value of $\gamma$ in the $\delta$ phase as compared to $\alpha$-Pu. Actual values of $\gamma$ obtained by means of
the LDA+FLEX technique are rather close to experimental results in both the $\alpha$ and $\gamma$ phases.

\begin{table}
\begin{tabular}{l|cc|cc|}
\hline
 &  \multicolumn{2}{c|}{ Effective mass } & \multicolumn{2}{c|}{$\gamma$ (mJ K$^{-2}$mol$^{-1}$)} \\
 &  $\alpha$-Pu & $\delta$-Pu &  $\alpha$-Pu & $\delta$-Pu \\
\hline 
LDA+FLEX \hm & \hm 3.85 &\hm  4.02 &\hm 20.7 &\hm 37.9  \hm \\
LDA+U+FLEX \hm & \hm 2.22 &\hm  1.68 &\hm 15.6 &\hm 8.6 \hm \\
Exper. \cite{Lash03,Lash05,havel06} & & & 17 & 43, 64, 35-55 \\
\hline
\end{tabular}
\caption{Effective mass $\frac{m^*}{m} $ and the  electronic 
specific heat coefficient $\gamma$}
\label{table_gam}
\end{table}

\begin{figure}
\includegraphics[width=0.40\textwidth, angle=270]{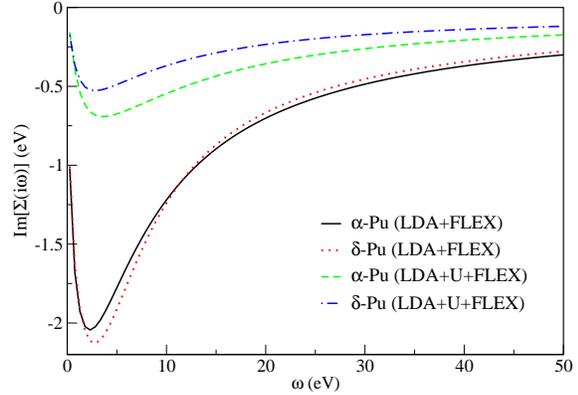}
\caption{The imaginary part of self-energy $\Sigma_{mm}(i\omega)$ for $m$=-3 on the Matsubara axis, 
computed for $\alpha$ and $\delta$-Pu
within the LDA+FLEX and LDA+U+FLEX techniques.} \label{pic_sig}
\end{figure}

In Fig.~\ref{pic_sig} we display the  self-energy $\Sigma_{mm}(i\omega)$ (for $m$=-3) computed 
on the Matsubara axis within the LDA+FLEX and LDA+U+FLEX techniques for $\alpha$ and $\delta$-Pu. One may notice
that $\Im \Sigma(i\omega)$ obtained by LDA+U+FLEX is about four times smaller than one calculated 
within LDA+FLEX. That should be expected on the basis of already observed tendency for $m^*$.
We stress again, that in both descriptions  (LDA+FLEX and LDA+U+FLEX ) we have a quasiparticle feature
in the spectra. This feature, 
becomes  the famous Kondo resonance  as  U increases.
However, the  values of the many body  renormalizations , namely the value of the slope   
of $\Im \Sigma(i\omega)$ obtained in the LDA+U+FLEX calculations of $\delta$-Pu 
is  relatively small 
and closer to the non interacting system (see Fig.~\ref{pic_sig})
while the peak at $E_F$ that is present in the    LDA+FLEX DOS is closer to the Fermi level and more enhanced.

\begin{figure}
\includegraphics[width=0.40\textwidth, angle=270]{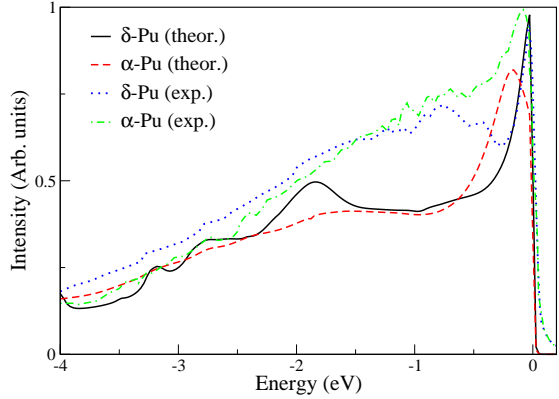}
\caption{The theoretical LDA+FLEX and experimental PES \cite{arko2000} for $\alpha$ and $\delta$-Pu} \label{pes_comp}
\end{figure}

Finally in Fig.~\ref{pes_comp} we compare the $\alpha$ and $\delta$-Pu DOS obtained within the "open-shell" 
LDA+FLEX approach to the experimental photoemission spectra, collected at a photon energy
of 40.8 eV, where the band 6$d$ and 5$f$ cross section are nearly equal \cite{arko2000}. In order to produce
the theoretical spectra we have multiplied the total DOS by the Fermi function, corresponding to experimental 
temperature of 80~K and normalized it to equal spectral weight  between -4 and 0 eV for $\alpha$ and $\delta$-Pu
as was done in analysis of the experimental PES in Ref.~\cite{arko2000}.  In result one has very good agreement between
experimental and theoretical PES in vicinity of the Fermi level. For lower energies below -0.5 eV the agreement is less
satisfactory, but there the $d$ states should give appreciated contribution to PES, therefore the accurate evaluation
of the corresponding 
6$d$ and 5$f$ matrix elements 
may become necessary for a faithful representation of PES. Moreover, the 
multiplet effects in the 5$f$ shell should be taken into account 
and these are beyond our simple  T-matrix FLEX impurity solver.

In conclusion, we have carried out comparative analysis of the specific heat and electronic structure in $\alpha$ and
$\delta$-plutonium by means of the LDA+DMFT technique in conjunction with the 
perturbative T-matrix and fluctuating exchange quantum impurity solver. We have shown that by assuming 
the "closed" $f^6$ configuration of the Pu $f$ shell one can not explain high value of the electronic 
specific heat coefficient $\gamma$ in $\delta$-Pu as well as substantial enhancement of $\gamma$ in the $\delta$
phase as compared to $\alpha$-Pu, even in the case when the dynamical fluctuations are properly taken into account.
At the same time, by assuming rather the "open shell" configuration, which is closer to $f^5$ one may obtain
the observed enhancement of  $\gamma$ in the $\delta$ phase and actual values of $\gamma$ in rather good agreement
with the experiment. The experimental PES is also well reproduced by the LDA+DMFT DOS in the "open shell" configuration.
This suggests that the actual \cite{Solovyev:1991}  ground state of the $f$ shell in Pu should be rather closer to the
"open shell" $f^5$ configuration than   $f^6$. 
This picture is consistent with the results of  a series of    EELS and XAS 
studies of Ref.~\cite{Tobin:2005} and 
in  good agreement with  the results
of recent  
LDA+DMFT \cite{Shim:2006} calculations, where more  sophisticated solvers were used 
to access the  strong correlation regime. 
The good agreement with respect to experiments achieved
by our work  suggest  the possibility of exploring 
the paramagnetic state of  inhomogeneous alloys and interfaces of actinides using
DMFT LDA+FLEX.
Notice  in the context 
that simplified solvers such as     the iterated perturbation theory was very useful
for the study of   the paramagnetic
phase of the one band Hubbard model, but far worse in the description of magnetically ordered phases.  
For this reason we have not addressed in this paper the issue of magnetic long range order.  
This issue  is  very delicate,   because
as is well known in the context of the Anderson impurity model, approximate  perturbative treatments that
signal the onset of magnetism, can indicate the need to include explicitly Kondo screening
rather  than the occurrence of a magnetic instability. 

\vspace{0.5cm}
\section*{ACKNOWLEDGMENTS}

We would like to thank A. Georges for useful discussions.
This work was funded by the  NNSA SSAA program. 
LP acknowledges support from CNRS, \'{E}cole Polytechnique and the E. U. "Psi-k f-electron"
Network under contract HPRN-CT-2002-00295.

\end{document}